\documentclass[aps,prl,twocolumn,superscriptaddress]{revtex4}
\usepackage{graphicx}
\usepackage{epstopdf}
\usepackage{color}

\begin{document}
%\preprint{}

\title{Linear scaling relation between two-dimensional massless Dirac fermion Fermi velocity and Fe-As bond length in iron arsenide superconductor systems}

\author{Chengpu Lv$^{\dagger}$}
\affiliation{Beijing National Laboratory for Condensed Matter Physics, Institute of Physics, Chinese Academy of Sciences, Beijing 100190, China}
\affiliation{School of Physical Sciences, University of Chinese Academy of Sciences, Beijing 100190, China}

\author{Jianzhou Zhao$^{\dagger}$}
\affiliation{Co-Innovation Center for New Energetic Materials, Southwest University of Science and Technology, Mianyang 621010 Sichuan, China}

\author{Yueshan Xu$^{\dagger}$}
\affiliation{Beijing National Laboratory for Condensed Matter Physics, Institute of Physics, Chinese Academy of Sciences, Beijing 100190, China}
\affiliation{School of Physical Sciences, University of Chinese Academy of Sciences, Beijing 100190, China}
	
\author{Yu Song}
\affiliation{Department of Physics and Astronomy, Rice University, Houston, Texas 77005, USA}

\author{Chenglin Zhang}
\affiliation{Department of Physics and Astronomy, Rice University, Houston, Texas 77005, USA}

\author{Mykhaylo Ozerov}
\affiliation{National High Magnetic Field Laboratory, Tallahassee, Florida 32310, USA}

\author{Pengcheng Dai}
\affiliation{Department of Physics and Astronomy, Rice University, Houston, Texas 77005, USA}

\author{Nan-Lin Wang}
\affiliation{International Center for Quantum Materials, School of Physics, Peking University, Beijing 100871, China}

\author{Zhi-Guo Chen}
\email{zgchen@iphy.ac.cn} 
\affiliation{Beijing National Laboratory for Condensed Matter Physics, Institute of Physics, Chinese Academy of Sciences, Beijing 100190, China}
\affiliation{Songshan Lake Materials Laboratory, Dongguan, Guangdong 523808, China}

\begin{abstract}
{  Two-dimensional (2D) massless Dirac fermions (MDF), which represent a type of quasi-particles with linear energy-momentum dispersions only in 2D momentum space, provide a fertile ground for realizing novel quantum phenomena. However, 2D MDF were seldom observed in the superconducting bulk states of 3D materials. Furthermore, as a cornerstone for accurately tuning the quantum phenomena based on 2D MDF, a quantitative relationship between 2D MDF and a structural parameter has rarely been revealed so far. Here, we report magneto-infrared spectroscopy studies of the iron-arsenide-superconductor systems NaFeAs and $A\mathrm{Fe_2As_2} (A = \mathrm{Ca,~Ba})$ at temperature $T \sim  4.2 $ K and at magnetic ﬁelds ($B$) up to 17.5 T. Our results demonstrate the existence of 2D MDF in the superconducting bulk state of NaFeAs. Moreover, the 2D-MDF Fermi velocities in NaFeAs and $A\mathrm{Fe_2As_2} (A = \mathrm{Ca, Ba})$, which are extracted from the slopes of the linear $\sqrt{B}$ dependences of the Landau-level transition energies, scale linearly with the Fe-As bond lengths. The linear scaling between the 2D-MDF Fermi velocities and the Fe-As bond lengths is supported by (i) the linear relationship between the square root of the eﬀective mass of the $d_{xy}$  electrons and the Fe-As bond length and (ii) the linear dependence of the square root of the calculated tight-binding hopping energy on the Fe-As bond length. Our results open up new avenues for exploring and tuning novel quantum phenomena based on 2D MDF in the superconducting bulk states of 3D materials.  }   

\end{abstract}

%\pacs{}

\maketitle

Two-dimensional (2D) massless Dirac fermions (MDF) in condensed matter have attracted enormous interest due to their crucial roles in achieving various exotic quantum phenomena, which include novel quantum Hall effect \cite{1,2,3}, Klein tunnelling \cite{4} and giant linear magneto-resistance \cite{5}. Noteworthily, a large part of quantum phenomena based on 2D MDF can be largely influenced by the 2D-MDF Fermi velocity \cite{6,7}. Thus, a scaling relation concerning 2D-MDF Fermi velocity can serve as an important basis for accurately tuning the related quantum phenomena. Previously, 2D-MDF Fermi velocity was mostly revealed to have the quantitative relationships with the physical quantities depicting electron-electron interactions \cite{6,7}. However, the scaling between 2D-MDF Fermi velocity and a quantity concerning the other degree of freedom remains elusive. Furthermore, most of 2D MDF were observed on the surfaces of 3D topological insulators and in 2D materials, such as graphene \cite{8,9,10,11,12}. Few 3D materials host 2D MDF in their bulk.

\begin{figure}[ht]
	\centering
	\includegraphics[width=0.47\textwidth]{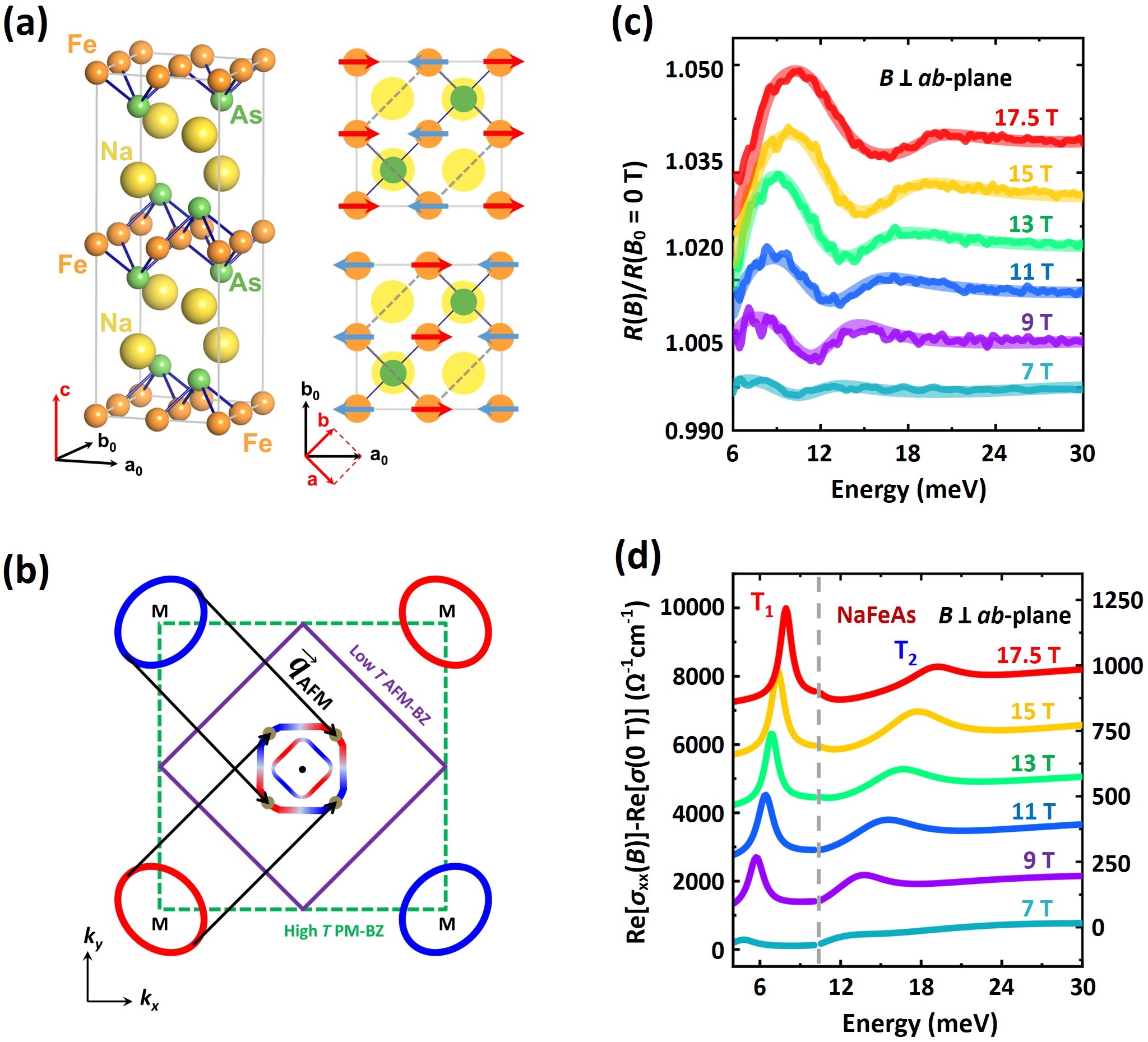}
	
	\caption{\label{Fig1} Antiferromagnetic (AFM) order, Fermi surfaces and magneto-optical response of NaFeAs. (a) Left: orthorhombic unit cell of NaFeAs in the AFM state, containing two FeAs layers. Right: collinear AFM order in two FeAs layers. The dashed and solid grey lines display the paramagnetic (PM) unit cell within the FeAs layer at high temperatures and the unit cell after the AFM phase transition, respectively. (b) Schematic of the Fermi surfaces of NaFeAs in the PM state. Four node (i.e. Dirac) points (grey dots) and Dirac cones emerge after the folding of the PM BZ along the AFM wave vector (the black arrows). (c) Relative reflectance spectra $R(B)/R(B_0= 0~\mathrm{T})$ of NaFeAs at different magnetic fields. The solid curves with noise and the partially transparent curves in (c) represent the measured $R(B)/R(B_0= 0~\mathrm{T})$ spectra of NaFeAs and the magneto-optical fits to the measured spectra, respectively. (d) Relative real part of the optical conductivity of NaFeAs. The $R(B)/R(B_0= 0~\mathrm{T})$ spectra in (c), the $\rm T_1$ peak and the $\rm T_2$ peak in (d) are displaced from one another by 0.009, 1400 ($\rm \Omega^{-1}cm^{-1}$) and 200 ($\rm \Omega^{-1}cm^{-1}$) for clarity, respectively.}
\end{figure}

Previous investigations indicated that in the antiferromagnetic (AFM) bulk states of 3D iron-arsenide-superconductor systems (see Fig.~\ref{Fig1}(a)) \cite{13,14}, the paramagnetic-Brillouin-zone folding results in the crossing points of the linearly dispersing bands, which are \textit{topologically protected} by the AFM order, inversion symmetry and a combination of time-reversal and spin-reversal symmetry, emerges near Fermi energy ($E_F$) (see Fig.~\ref{Fig1}(b)) \cite{15,16,17,18,19,20,21,22,23,24,25,26,27,28,29,30,31}. In addition, angle-resolved photoemission spectroscopy (ARPES) studies showed that linear band dispersions within the $k_x-k_y$ plane are present below $E_F$ in the low-temperature AFM bulk state of iron-arsenide-superconductor systems \cite{17,23,27,28}. Therefore, MDF with linear energy-momentum dispersions and topologically protected band-crossing points are present in the AFM bulk states of iron-arsenide-superconductor systems \cite{15,16,17,18,19,20,21,22,24,25,26}. However, to date, only “122”-type iron arsenides have been identified both experimentally and theoretically to possess topologically nontrivial 2D MDF in their AFM bulk \cite{29}. 2D MDF have yet been observed in the superconducting bulk states of 3D iron-arsenide-superconductor systems. A natural question to ask is whether a broader system of iron-arsenide superconductors can be found to have topologically nontrivial 2D MDF in the bulk state. Moreover, a scaling relationship about the 2D-MDF Fermi velocity in iron-arsenide-superconductor systems, which is expected to be significant for tuning exotic quantum phenomena based on their 2D MDF, is still lacking.

Topologically nontrivial 2D MDF have been discovered on the surfaces of iron-based superconductors \cite{32,33,34}. More importantly, owing to the proximity effect from the superconducting bulk, a superconducting gap would be opened in the topologically nontrivial 2D MDF state, which can induce topological superconductivity on the surfaces of iron-based superconductors \cite{35,36,37,38}. As a key feature of topological superconductivity, Majorana zero modes have been reported to emerge inside the vortex cores of the superconducting surfaces of iron-based superconductors, which offers a potential platform for realizing topological quantum computing \cite{39,40,41,42,43}. It is worth noticing that at ambient pressure, 3D iron-arsenide superconductor NaFeAs with the superconducting transition temperature $T_c \sim{} 23 ~ \mathrm{K}$ exhibits an AFM order below 39 K, which (i) means that an AFM order and superconductivity coexist in the bulk ground state of NaFeAs and (ii) implies that 2D MDF may be present in its superconducting bulk state. More importantly, searching for topologically nontrivial 2D MDF in the superconducting bulk state of NaFeAs is expected to provide a crucial clue to exploring novel quantum phenomena including topological superconductivity. Nonetheless, whether 2D MDF exist in the superconducting bulk state of NaFeAs is still unclear.

Magneto-infrared spectroscopy is a bulk-sensitive experimental technique for investigating 2D MDF in materials as it can probe the Landau level (LL) transitions of 2D MDF. Here, we performed the magneto-infrared spectroscopy measurements of the single crystals of NaFeAs and $\rm CaFe_2As_2$ at  $T\sim$ 4.2 K with the applied magnetic field parallel to the wave vector of the incident light and the crystalline $c$-axis (see the details about the single-crystal growth and magneto-infrared measurements in Supplementary Materials). Optical absorption features arising from the LL transitions were observed in NaFeAs and $\rm CaFe_2As_2$. Our observation of the $\sqrt{B}$ dependences of the measured LL transition energies, the zero-energy intercepts at $B = 0 \rm~ T$ under the linear extrapolations of the LL transition energies, the energy ratio ($\sim$ 2.4) between the two observed LL transitions, and the dominance of the zeroth-LL-associated optical absorption features, combined with the DFT+DMFT-calculation-derived band dispersions which are linear within the 2D $k_x-k_y$ plane in momentum space, but quite weak along the $k_z$ direction, indicates the existence of 2D MDF in the superconducting and AFM bulk state of NaFeAs. Furthermore, the LL transition energy of 2D MDF in AFM $\rm CaFe_2As_2$ also shows the linear dependence on $\sqrt{B}$ and has the zero-energy intercept at $B$ = 0 T under the linear extrapolations. In addition, the 2D-MDF Fermi velocities in NaFeAs, $\rm BaFe_2As_2$ and $\rm CaFe_2As_2$, which are obtained from the slopes of the linear $\sqrt{B}$ dependences of the LL transition energies, is linear with the Fe-As bond length—a quantity about lattice degree of freedom, which is supported by not only the linear relationship between the square root of the effective mass of the $d_{xy}$ electrons and Fe-As bond length but also the linear scaling between the square root of the calculated tight-binding hopping energy and Fe-As bond length.

\begin{figure*}[ht]
	\centering
	\includegraphics[width=0.94\textwidth]{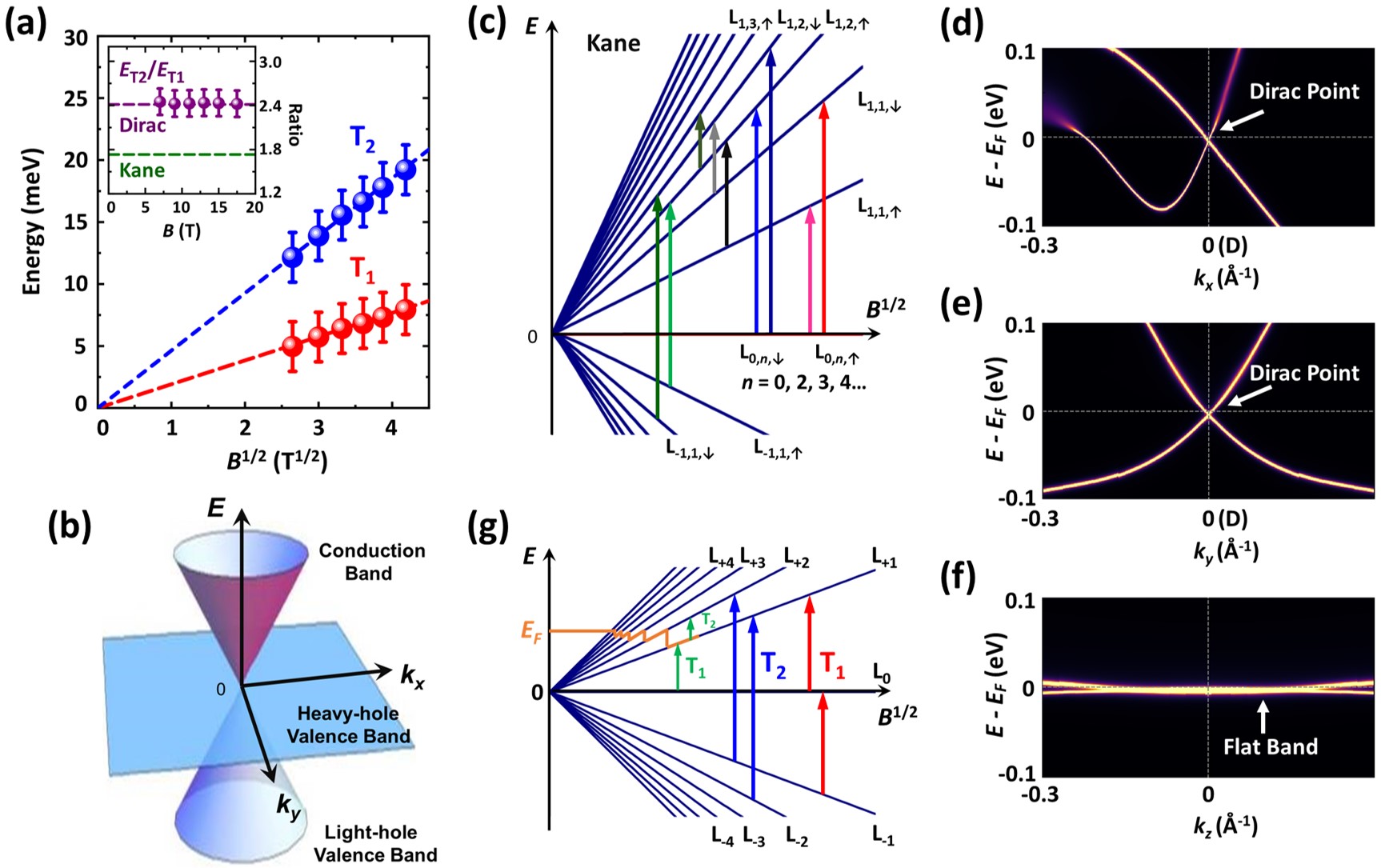}
	%	\subfigure{
		%		\includegraphics[width=0.47\textwidth]{figure/vertical R ?}}
	%	\subfigure{
		%		\label{Fig.sub.2}
		%			\includegraphics[width=0.23\textwidth, angle=0]{figure/$\sigma_1$($\omega$)}}
	\caption{\label{Fig2} Landau level transitions and band dispersions of NaFeAs in the AFM state. (a) Observed two LL transitions, $\rm T_1$ and $\rm T_2$, in an $E-\sqrt{B}$ plot. The inset shows energy ratio of the two LL transitions as a function of $B$. The purple and green dashed lines show the theoretical energy ratios which are based on massless Dirac and Kane model, respectively. (b) Band dispersions of massless Kane fermions. (c) Landau level spectrum of massless Kane fermions with $k_z  = 0$. The allowed optical transitions are indicated by the arrows. (d)-(f) Calculated band dispersions of NaFeAs along $k_x$, $k_y$ and $k_z$ near $E_F$. The flat band in (f) represents the energy-momentum dispersion of MDF along $k_z$ in NaFeAs. (g) Schematic of the LL transitions of 2D MDF in NaFeAs. The magnetic field dependence of the assumed Fermi level is shown by the orange lines. The two green arrows indicate the assumed $\rm T_1$ and $\rm T_2$ in case (i).}
\end{figure*}

Fig.~\ref{Fig1}(c) displays the relative reflectance spectra $R(B)/R(B_0  = 0~\mathrm{T})$  of NaFeAs measured at different magnetic fields. Peak-like features are present in the $R(B)/R(B_0  = 0~\mathrm{T})$  and systematically move towards higher energies as the magnetic field is enhanced. To identify the nature of the peak-like features in the $R(B)/R(B_0  = 0~\mathrm{T})$ , we obtained the real part of the diagonal optical conductivity $\mathrm{Re}\left[\sigma_{xx} (B,\omega) \right]$ by fitting the absolute reflectance spectra $R(B)$ based on the magneto-optical model (see the details about the fitting based on the magneto-optical model in Supplementary Materials) \cite{51}. In Fig.~\ref{Fig1}(d), the peak-like features can be observed in the relative optical conductivity spectra $\mathrm{Re}\left[\sigma_{xx} (B,\omega) \right] - \mathrm{Re}\left[\sigma(B = 0~\mathrm{T},\omega)\right]$ of NaFeAs. Considering that (i) the highest magnetic field ($B_{max}  = 17.5~\mathrm{T}$) applied along the crystalline $c$-axis here is much lower than that ($B > 500$ T) for dramatically destroying the AFM order in iron-arsenide-superconductor systems at low temperatures \cite{52}, and (ii) the energy region ($\sim$ 30 meV) where the peak-like features are present in Fig.~\ref{Fig1}(d) is narrower than the energy region ($\sim$ 100 meV) where the spectral weight is transferred due to the AFM phase transition in NaFeAs \cite{53}, the observed peak-like features in the $R(B)/R(B_0  = 0~\mathrm{T})$ and the $\mathrm{Re}\left[\sigma_{xx} (B,\omega) \right] - \mathrm{Re}\left[\sigma(B = 0~\mathrm{T},\omega)\right]$ here are irrelevant with the magnetic-fied-induced change in the AFM order of NaFeAs. Furthermore, the direction of the magnetic field here is perpendicular to that (orthorhombic $a$- or $b$-axis) of the magnetic field used for partially detwinning the single crystals of iron-arsenide superconductor systems \cite{54,55,56}, so the peak-like features in Figs.~\ref{Fig1}(c) and \ref{Fig1}(d) are unlikely to arise from the detwinning.

To further study the origin of the observed peak-like features in the superconducting and AFM state of NaFeAs, we plotted the energies of the peak-like features in the $\mathrm{Re}\left[\sigma_{xx} (B,\omega) \right] - \mathrm{Re}\left[\sigma(B = 0~\mathrm{T},\omega)\right]$ spectra as a function of $\sqrt{B}$ in Fig.~\ref{Fig2}(a), which shows that the energies (i.e., $E_{\rm T1}$ and $E_{\rm T2}$) of the two peak-like features $\rm T_1$ and $\rm T_2$ not only exhibit a linear dependence on $\sqrt{B}$ and the zero intercepts at $B =0~\rm T$ under linear extrapolations but also have the ratio $E_\mathrm{T2}/E_\mathrm{T1}\approx 2.4$. Therein, the  $\sqrt{B}$ dependences of the $\rm T_1$ and $\rm T_2$ energies provide a signature of the fermions with linear energy-momentum dispersions. Moreover, the zero intercepts at $B = 0~\rm T$ under the linear extrapolations of the $\rm T_1$ and $\rm T_2$ energies indicate that the energy gap between the linearly dispersing valence and conduction bands is zero (i.e., the presence of band-degeneracy points) and that the corresponding fermions have zero mass. Thus, the $\sqrt{B}$-dependence of the peak energy positions transition energies and the zero-energy intercepts at $B = 0~\rm T$ manifest the existence of massless fermions with linear dispersions and band-degeneracy points in the superconducting and AFM bulk state of NaFeAs.

In condensed matter, there are mainly three kinds of massless fermions with linear dispersions and band-degeneracy points, which include MDF, Kane fermions, and Weyl fermions \cite{1,2,3,57,58,59,60,61,62}. In 3D momentum space, the Kane fermions with zero bandgap have a flat band around the band-degeneracy point and two linearly dispersing bands—light conduction band and light valence band with the linear energy-momentum dispersions $E_k= \pm \hbar v |k|$, where $k$ is the 3D crystal momentum, and $v$ is the Fermi velocity (see Fig.~\ref{Fig2}(b)) \cite{57,58}. The LL spectrum of the Kane fermions, which includes the spin splitting of LLs \cite{57,58}, has the form:

	\begin{eqnarray}
	E_{\xi,n,\sigma}^{Kane} = v_F^K \xi \sqrt{e \hbar B \left(2n - 1 + \frac{\sigma}{2}\right) + \hbar^2 k_z^2},
	\label{Eq1}
	\end{eqnarray}

where $v_F^K$ is the effective Fermi velocity of the LLs of the Kane fermions, the band index $\xi = -1,0,+1$ for the light-hole valence band, the flat band and the light conduction band, respectively, the spin index $\sigma = \rm +1~and~-1$ for spin-down $\downarrow$ and spin-up $\uparrow$, respectively, and the magnetic field is along $z$ direction (i.e. crystalline $c$-axis).  Considering that (i) the optical transitions between the LLs (from $\mathrm{LL}_n$ to  $\mathrm{LL}_n'$ ) of massless Kane fermions obey the selection rule: $\Delta n = n - n’ = \pm 1, \Delta \sigma = 0 $ and no restriction on $\xi$ (see Fig.~\ref{Fig2}(c)) \cite{57}, and (ii) the magneto-optical absorption of massless Kane fermions is mainly contributed by the LL transitions at $k_z  = 0$, the two \textit{inter}-LL transitions, $ \mathrm{LL}_{n=0,\xi=0,\downarrow}  \rightarrow \mathrm{LL}_{n=1,\xi=1,\downarrow}$ and $\mathrm{LL}_{n=0,\xi=0,\uparrow}  \rightarrow \mathrm{LL}_{n=1,\xi=1,\uparrow}$ (or $ \mathrm{LL}_{n=0,\xi=2,\downarrow}  \rightarrow \mathrm{LL}_{n=1,\xi=1,\downarrow}$ and $ \mathrm{LL}_{n=0,\xi=2,\uparrow}  \rightarrow \mathrm{LL}_{n=1,\xi=1,\uparrow}$) have the maximal energy ratio among the adjacent LLs: $\sqrt{3/2}: \sqrt{1/2}  \approx 1.732$, which is distinctly \textit{smaller} than that between the observed $\rm T_2$ and $\rm T_1$ (see the inset of Fig.~\ref{Fig2}(a)). Additionally, previous theoretical studies showed that the flat band of Kane fermions is absent around the band-degeneracy point of the linearly dispersing bands of iron-arsenide-superconductor systems \cite{15,16}. Thus, the observed peak-like features in the $R(B)/R(B_0  = 0 \rm~T)$ and the $\mathrm{Re}\left[\sigma_{xx} (B,\omega) \right] - \mathrm{Re}\left[\sigma(B = 0~\mathrm{T},\omega)\right]$ spectra of NaFeAs are very unlikely to be contributed by the optical transitions between the LLs of Kane fermions.

Weyl fermions are characterized by the linear dispersions in three-dimensional (3D) momentum space and pairs of degenerate Weyl points with opposite chirality \cite{59,60,61,62}. Since (i) the zeroth LLs of Weyl fermions with 3D linear band dispersions do not have a singularity of the density of states (DOS), and (ii) the optical absorptions are determined by the joint DOS of the initial and final states, the lowest-energy peak-like feature arising from the zeroth-LL-related inter-LL transition $\rm LL_{-1}  \rightarrow LL_{\pm0}$ (or $ \rm LL_{\pm0}  \rightarrow LL_{+1}$) in the $ \mathrm{Re}[\sigma_{xx} (B)] - \mathrm{Re}[\sigma(0 ~\rm T)]$ spectra of Weyl fermions should be \textit{weaker} than the second lowest-energy peak-like feature induced by the inter-LL transition $\rm LL_{-1}  \rightarrow LL_{+2} $ or $\rm (LL_{-2 } \rightarrow LL_{+1})$ \cite{63,64}, which is on the contrary to the case that the intensity of the lowest-energy peak-like feature $\rm T_1$ in the relative optical conductivity spectra of NaFeAs shown in Fig.~\ref{Fig1}(d) is much higher than that of the second lowest energy peak-like feature $\rm T_2$. Besides, previous theoretical studies showed that the degenerate nodal points in iron-arsenide-superconductor systems have the same chirality \cite{15,16}, which is in contrast to the opposite chirality of a pair of Weyl nodes \cite{59,60,61,62}. Therefore, the dominance of the zeroth-LL-related absorption features in the $ \mathrm{Re}[\sigma_{xx} (B)] - \mathrm{Re}[\sigma(0 ~\rm T)]$ spectra of NaFeAs, together with the same chirality of the degenerate nodal points in iron-arsenide-superconductor systems, indicates that the observed peak-like features in Figs.~\ref{Fig1}(c) and \ref{Fig1}(d) should be irrelevant with the optical transitions between the LLs of Weyl fermions. Based on the above discussion, we assign the peak-like features in the magneto-optical spectra of NaFeAs to the LL transitions of MDF.

For 2D MDF, the lowest-energy the peak feature of the zeroth-LL-related inter-LL transition $\rm LL_{-1 } \rightarrow LL_{\pm 0}$ (or $\rm LL_\pm 0)  \rightarrow LL_{+1}$) in $ \mathrm{Re}[\sigma_{xx} (B)] - \mathrm{Re}[\sigma(0 ~\rm T)]$ is stronger than the second lowest-energy peak feature of the inter-LL transition $ \rm LL_{-1}  \rightarrow LL_{+2}$ (or $ \rm LL_{-2}  \rightarrow LL_{+1}$) because of a DOS singularity of the zeroth LLs of 2D Dirac fermions \cite{29,63,64}, which is consistent with the relative intensity between the peak-like features $\rm T_1$ and $\rm T_2$ in Fig.~\ref{Fig1}(d), while for 3D MDF, the lowest-energy peak feature of the zeroth-LL related inter-LL transition $\rm LL_{-1} \rightarrow LL_\pm 0$ (or $\rm LL_{\pm 0}  \rightarrow LL_{+1}$) in $\mathrm{Re}[\sigma_{xx} (B)] - \mathrm{Re}[\sigma (0~ \rm T)]$ is \textit{weaker} than the second lowest-energy peak feature of the inter-LL transition $ \rm LL_{-1 } \rightarrow LL_{+2}$ or ($\rm LL_{-2}  \rightarrow LL_{+1}$) due to the absence of a DOS singularity of the zeroth LLs of 3D MDF \cite{65}, which is similar to the case for Weyl fermions. Thus, the dominance of the peak-like feature $\rm T_1$ in the $ \mathrm{Re}[\sigma_{xx} (B)] - \mathrm{Re}[\sigma(0 ~\rm T)]$ indicates that MDF in the superconducting and AFM bulk state of NaFeAs should be 2D. Besides, in Figs.~\ref{Fig2}(d) to \ref{Fig2}(f), our DFT+DMFT calculations show that in the superconducting and AFM state of NaFeAs, the bands near $E_F$ disperse linearly along the $k_x$ and $k_y$ directions, but disperse quite weakly along the $k_z$  direction (see the details about the calculations in Supplementary Materials), which further supports 2D MDF in this iron-arsenide superconductor.

The LL spectrum of 2D MDF without considering Zeeman effects can be given by \cite{29,49,50}:
	\begin{eqnarray}
    E_{n}^{2D} = \text{sgn}(n) \, v_{F}^{D} \sqrt{2 e \hbar |n| B},
    \label{Eq2}
    \end{eqnarray}
where $v_{F}^{D}$ is the 2D-MDF Fermi velocity, the integer $n$ is LL index, $\mathrm{sgn}(n)$ is the sign function, $e$ is the elementary charge and $\hbar$ is Planck’s constant divided by $2\pi$. Given the selection rule for the allowed optical transitions from $\rm LL_n $ to  $\rm LL_{n'}$ \cite{29,49,50}:
	\begin{eqnarray}
    \Delta n = |n| - |n'| = \pm 1,
    \label{Eq3}
    \end{eqnarray}
the energies $E_{n \to n'}^{2D}$ of the allowed optical transitions from $\rm LL_n $ to  $\rm LL_{n'}$ have the form:
	\begin{eqnarray}
    E_{n \to n'}^{2D} = v_F^D \sqrt{B} \left[ \text{sgn}(n') \sqrt{2 e \hbar |n'|} - \text{sgn}(n) \sqrt{2 e \hbar |n|} \right].
    \label{Eq4}
    \end{eqnarray}

According to Eqs.~(\ref{Eq2}) and (\ref{Eq4}), two groups of LL transitions of 2D MDF have the energy ratios which are close to that ($\sim 2.4$)  between $\rm T_2$ and $\rm T_1$: (i) the energy ratio $1:(\sqrt{2}-1) \approx 2.414$  between the LL transitions, $\rm LL_{-1} \rightarrow LL_0$ and $\rm LL_{-2} \rightarrow LL_{-1}$ (or $\rm LL_0 \rightarrow LL_{+1}$ and $\rm LL_{+1} \rightarrow LL_{+2} $), and (ii) the energy ratio $(1+\sqrt{2}): 1 \approx 2.414$ between the LL transitions, $\rm LL_{-1} \rightarrow LL_{+2}$ and $\rm LL_{-1} \rightarrow LL_0 $(or $\rm LL_{-2} \rightarrow LL_{+1} $ and $\rm LL_0 \rightarrow LL_{+1}$). To extract the 2D-MDF Fermi velocity in NaFeAs, it is essential to determine which group of LL transitions of 2D MDF leads to the two peak-like features $\rm T_1$ and $\rm T_2$. If we assumed that the peak-like features $\rm T_1$ and $\rm T_2$ separately arise from the LL transitions $\rm LL_{-2} \rightarrow LL_{-1}$ and $\rm LL_{-1} \rightarrow LL_{0}$  (or $\rm LL_{+1} \rightarrow LL_{+2}$ and $\rm LL_{0} \rightarrow LL_{+1}$) (see the two green arrows in Fig.~\ref{Fig2}(g)), the $E_F$ should be still pinned on $\rm LL_{+1}$ or $\rm LL_{-1}$ at $B=17.5 ~\rm T$ due to the presence of of the intra-LL transition $\rm LL_{-2} \rightarrow LL_{-1}$ (or $\rm LL_{+1} \rightarrow LL_{+2}$). Considering (i) the 2D-MDF Fermi velocity obtained based on Eq.~(\ref{Eq4}), $v_F^D \approx 1.30 \times 10^5$ m/s, and (ii) the $\rm LL_{+1}$ (or $\rm LL_{-1}$) energy calculated at $B=17.5~ \rm T$ based on Eq.~(\ref{Eq2}), the corresponding $E_F$ in NaFeAs is about 81 meV, which is much higher than the experimental $E_F$ of $\sim$ 2 meV measured by ARPES and the theoretical $E_F$ of $\sim$ 3 meV gotten by our DFT+DMFT calculations. Moreover, as the magnetic field increases, electrons in $\rm LL_{+1}$ (or holes in $\rm LL_{-1}$) would be depopulated, which would lead to the decrease in the spectral weight of the lowest-energy absorption feature resulting from the assumed LL transition $\rm LL_{-2} \rightarrow LL_{-1}$ (or $\rm LL_{+1} \rightarrow LL_{+2}$). In contrast, the lowest-energy peak-like feature $\rm T_1$ in the $\mathrm{Re}\left[\sigma_{xx} (B,\omega) \right] - \mathrm{Re}\left[\sigma(B = 0~\mathrm{T},\omega)\right]$ of NaFeAs becomes more dominant with increasing the magnetic field. Thus, the measured peak-like features $\rm T_1$ and $\rm T_2$ should not arise from the LL transitions in case (i): $\rm LL_{-1} \rightarrow LL_0$ and $\rm LL_{-2} \rightarrow LL_{-1}$ (or $\rm LL_0 \rightarrow LL_{+1}$ and $\rm LL_{+1} \rightarrow LL_{+2}$). For case (ii), the observation of the peak-like feature $\rm T_1$ which is assumed to come from the LL transition $\rm LL_{-1} \rightarrow LL_0$ (or $\rm LL_0 \rightarrow LL_{+1}$) at $B \le 9 $ T suggests that NaFeAs enters the quantum limit at low magnetic fields \cite{66,67}, which is in agreement with the low $E_F$ measured by ARPES \cite{24}. Moreover, in case (ii), based on Eq.~(\ref{Eq4}), the obtained 2D-MDF Fermi velocity $v_{F}^{D} \approx 5.4 \times 10^4$ m/s, which is comparable to the values deduced from the ARPES results. Therefore, the peak-like features $\rm T_1$ and $\rm T_2$ in the $\mathrm{Re}\left[\sigma_{xx} (B,\omega) \right] - \mathrm{Re}\left[\sigma(B = 0~\mathrm{T},\omega)\right]$ of NaFeAs should be assigned to the LL transitions, $\rm LL_{-1} \rightarrow LL_{+2}$ and $\rm LL_{-1} \rightarrow LL_0$ (or $\rm LL_{-2} \rightarrow LL_{+1}$ and $\rm LL_0 \rightarrow LL_{+1}$), respectively, which corresponds to the $v_{F}^{D} \approx 5.4 \times 10^4$ m/s.

\begin{figure}[ht]
	\centering
	\includegraphics[width=0.47\textwidth]{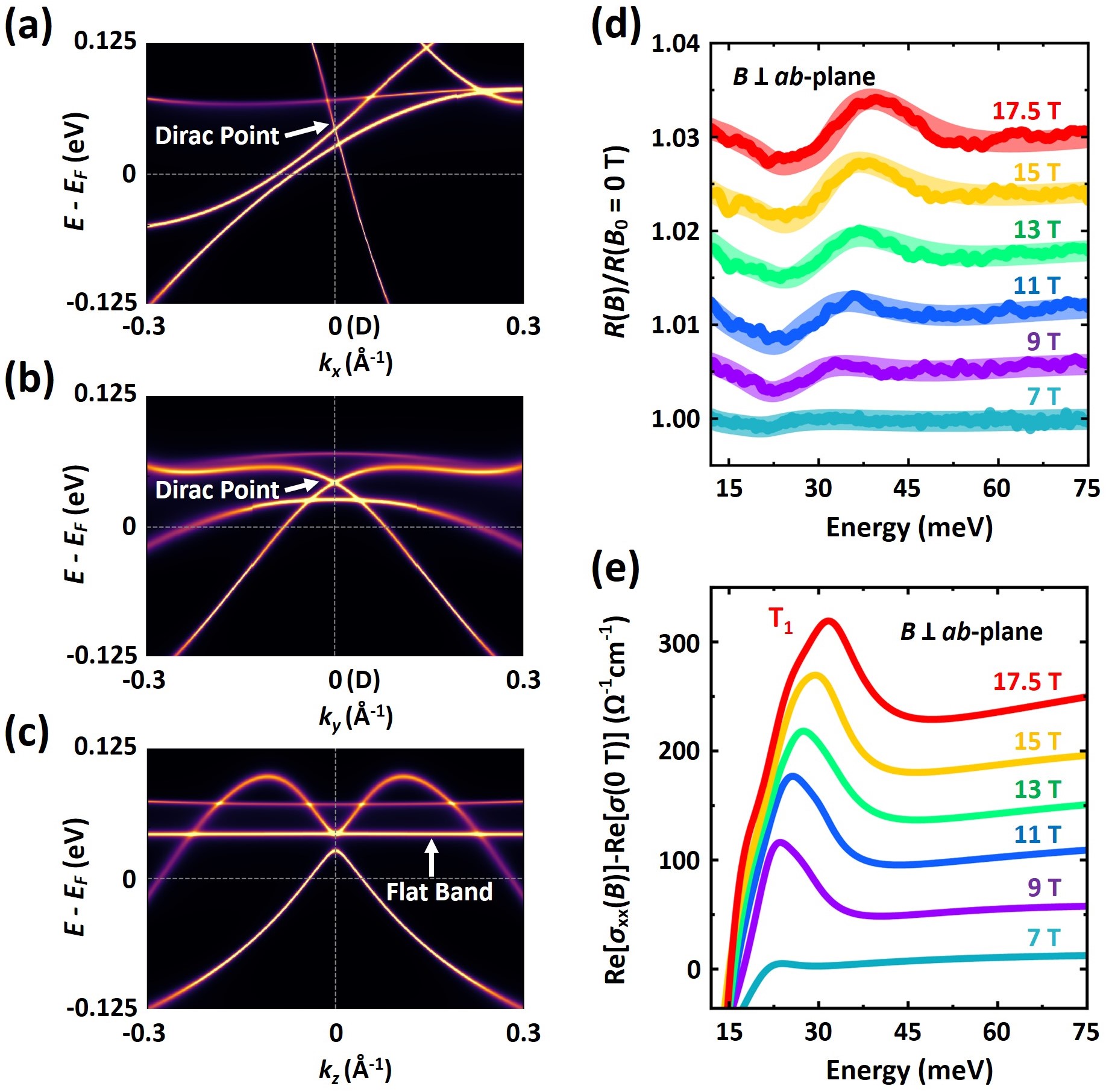}
	
	\caption{\label{Fig3}  Magneto-optical response and band dispersions of $\rm CaFe_2As_2$ in the AFM state. (a)-(c) Calculated energy-momentum dispersions of MDF in $\rm CaFe_2As_2$ along $k_x$, $k_y$ and $k_z$. The flat band in (c) represents the energy-momentum dispersion of MDF along $k_z$ in $\rm CaFe_2As_2$. (d) Relative reflectance spectra $R(B)/R(B_0= 0\rm ~ T)$ of $\rm CaFe_2As_2$. The solid curves with noise and the partially transparent curves in (d) represent the measured $R(B)/R(B_0= 0~\mathrm{T})$ spectra of $\rm CaFe_2As_2$ and the magneto-optical fits to the measured spectra, respectively. (e) Relative optical conductivity spectra of $\rm CaFe_2As_2$. The spectra in (d) and (e) are separately displaced from one another by 0.006 and 50 ($\rm  \Omega^{-1} cm^{-1}$) for clarity.}
\end{figure}

\begin{figure}[ht]
	\centering
	\includegraphics[width=0.47\textwidth]{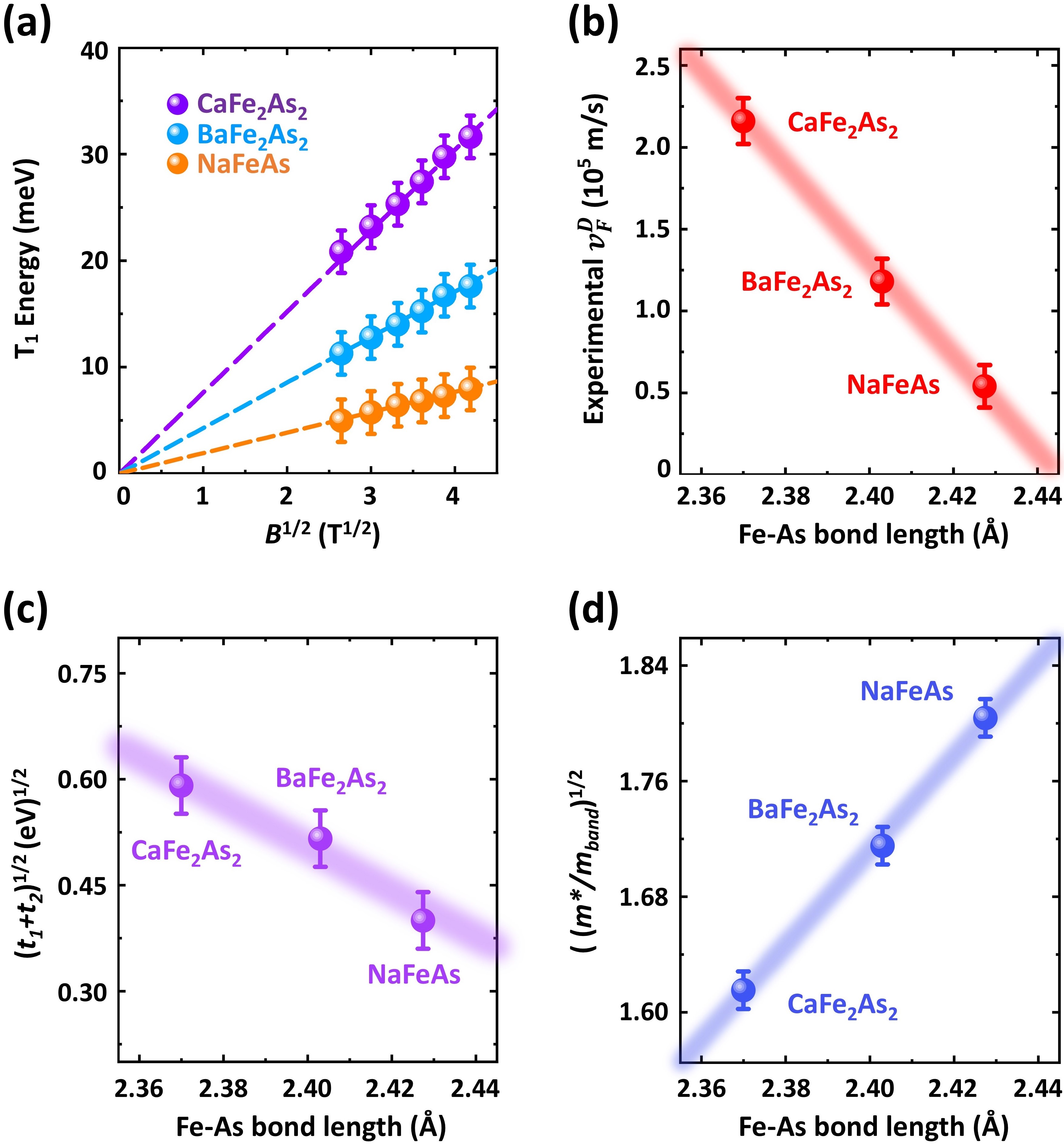}
	
	\caption{\label{Fig4}  Relationship between the 2D-MDF Fermi velocities and the Fe-As bond lengths in NaFeAs, $\rm BaFe_2As_2$ and $\rm CaFe_2As_2$. (a) $\sqrt{B}$ dependences of the $\rm T_1$ energies of NaFeAs, $\rm BaFe_2As_2$ and $\rm CaFe_2As_2$. (b) Linear scaling between the 2D-MDF Fermi velocities and the Fe-As bond lengths. (c) Linear relationship between $\sqrt{t_1 (xy,xy) + t_2 (xy,xy)}$ and the Fe-As bond lengths \cite{48}. (d) Linear scaling between $\sqrt{m^*/m_{band}}$ and the Fe-As bond lengths \cite{48}. $m_{band}$ here is the band mass.}
\end{figure}

To check whether 2D MDF exist in the bulk state of iron arsenide $\rm CaFe_2As_2$, we performed the DFT+DMFT calculations on its electronic bands (see the details about the calculations in Supplementary Materials). In Figs.~\ref{Fig3}(a) to \ref{Fig3}(c), our DFT+DMFT calculations exhibit the linear band dispersions along $k_x$ and $k_y$ directions and the very weak dispersions along $k_z$ direction, which indicates that $\rm CaFe_2As_2$ possesses 2D MDF as well. To obtain the 2D-MDF Fermi velocity in $\rm CaFe_2As_2$, we measured its relative reflectance spectra $R(B)/R(B_0  = 0 \rm ~T)$ at different magnetic fields (see the clear peak-like feature in each $R(B)/R(B_0  = 0 \rm ~T)$ spectrum in Fig.~\ref{Fig3}(d) and each $R(B)/R(B_0  = 0 \rm ~T)$ spectrum down to 6 meV in Fig. S4) and got its $\mathrm{Re}\left[\sigma_{xx} (B,\omega) \right] - \mathrm{Re}\left[\sigma(B = 0~\mathrm{T},\omega)\right]$ spectra via fitting the $R(B)$ (see Fig.~\ref{Fig3}(e) and the fitting parameters in Supplementary Materials). The energy region ($\sim$ 45 meV) where the peak-like features are present in the relative optical conductivity spectra of $\rm CaFe_2As_2$ in Fig. 3(e) is narrower than not only the energy region (up to at least 207 meV) where the spectral weight is transferred due to its AFM phase transition but also the energy ($\sim$ 82 meV) of its small spin-density-wave gap \cite{68}, which indicates that the observed peak-like features in the relative reflectance spectra and the relative optical conductivity spectra of $\rm CaFe_2As_2$ are unlikely to be relevant with the magnetic-field-induced change in its AFM order. Moreover, the energy position of the peak-like feature in Fig.~\ref{Fig4}(a) shows a linear dependence on $\sqrt{B}$ and has the zero-energy intercept at $B = 0 \rm ~T$ under the linear extrapolation as well, which suggests that the peak-like feature arises from the LL transition of 2D MDF. Given that (i) the $E_F$ obtained by our DFT+DMFT calculations is low and (ii) the spectral weight of the peak-like feature is enhanced with the growth of the magnetic field, the peak-like feature in Fig.~\ref{Fig3}(d) can be ascribed to the lowest-energy LL transition $\rm LL_{-1} \rightarrow LL_0$ (or $\rm LL_0  \rightarrow  LL_{+1}$) of 2D MDF. Then, we fitted the linear $\sqrt{B}$ dependence of the peak energy position based on Eq.~(\ref{Eq4}) and obtained the derived 2D-MDF Fermi velocity $v_F^D  \approx 2.16 \times 10^5$ m/s. The 2D-MDF Fermi velocity in $\rm CaFe_2As_2$ is larger than those in NaFeAs and $\rm BaFe_2As_2$ ($v_F^D \approx 1.18 \times 10^4$ m/s) \cite{29}, which is consistent with the fact that the peak-like feature in the relative optical conductivity spectra of $\rm CaFe_2As_2$ at $B = 17.5~\rm T$ is located at the highest energy position among the those of NaFeAs and $A\mathrm{Fe_2As_2}(A = \rm{Ba},~Ca)$ here.

The increase in the 2D-MDF Fermi velocity from NaFeAs to $A\mathrm{Fe_2As_2}(A = \rm{Ba}, Ca)$ inspires us to search for the relation between the 2D-MDF Fermi velocities and lattice parameters. Thus, we plotted the Fe-As-distance and As-Fe-As-angle dependences of the 2D-MDF Fermi velocities of NaFeAs, $\rm BaFe_2As_2$ and $\rm CaFe_2As_2$, respectively (see Fig.~\ref{Fig4}(b), Fig. S2, and the Fe-As distance and the As-Fe-As angle in Fig. 2(a) of Ref. \cite{48}). The quantitative relation between the 2D-MDF Fermi velocities and the As-Fe-As angles seems to be elusive (see Fig. S2). However, interestingly, the 2D-MDF Fermi velocities in NaFeAs, $\rm BaFe_2As_2$ and $\rm CaFe_2As_2$ scale linearly with the Fe-As bond lengths which are the quantities about lattice degree of freedom (see Fig. 4(b)). Previous DFT+DMFT calculations showed that the Dirac cones of iron-arsenide-superconductor systems near $E_F$ are mainly composed of iron $d_{xy}$ orbital \cite{69}. For iron-arsenide-superconductor systems, the shorter Fe-As distance suggests more overlap between the Fe-3$d$ and As-4$p$ orbitals, which means an easier hopping of the $d_{xy}$ electrons occupying the Dirac cone and thus implies a larger bandwidth ($W$). For 2D MDF with linear dispersions in iron-arsenide-superconductor systems, the larger bandwidth within the same momentum range ($\Delta k$) corresponds to the higher 2D-MDF Fermi velocity due to the relationship $v_F^D  = W/\Delta k$.

To check the linear scaling between the 2D-MDF Fermi velocities and the Fe-As bond lengths in NaFeAs, $\rm BaFe_2As_2$ and $\rm CaFe_2As_2$, we further studied the kinetic energy ($E_k$) and the effective mass ($m^*$) of the $d_{xy}$ electrons because the Dirac cones near $E_F$ are dominated by iron $d_{xy}$ orbital. In a low-energy effective tight-binding model, the hopping parameters ($t$) can be approximately regarded to be proportional to the kinetic energy $E_k$, so we plotted the square root of the sum of the DFT+DMFT-calculation-derived tight-binding hopping parameters ($t_1 (xy,xy)$ and $t_2 (xy,xy)$) for the electrons on the iron $d_{xy}$ orbital hopping to the orbital $d_{xy}$ of their nearest neighbor and next nearest neighbor iron atoms, i.e., $\sqrt{t_1 (xy,xy) + t_2 (xy,xy)}$ for NaFeAs, $\rm BaFe_2As_2$ and $\rm CaFe_2As_2$ with different Fe-As bond lengths in Fig.~\ref{Fig4}(c) (see the $t_1 (xy,xy)$ and $ t_2 (xy,xy)$ in Fig. S5 of the Supplementary Materials of Ref. \cite{48}). The $\sqrt{t_1 (xy,xy) + t_2 (xy,xy)}$ shows a linear relationship with the Fe-As bond length. The square root of the effective mass $m^*$ of the $d_{xy}$ electrons in Fig.~\ref{Fig1}(b) of Ref. \cite{48} exhibits a linear dependence on the Fe-As bond length as well (see Fig.~\ref{Fig4}(d)). Considering the linear relationship for “relativistic” fermions in solids $v_F^D  = \sqrt{(E_k/m^*)}$, the 2D-MDF Fermi velocities in NaFeAs, $\rm BaFe_2As_2$ and $\rm CaFe_2As_2$ should scale linearly with the Fe-As bond lengths.

In summary, our observation of the $\sqrt{B}$-dependence of the LL transition energies, the zero-energy intercept at $B = 0\rm ~T$ under linear extrapolations of the transition energies, the energy ratio ($\sim$ 2.4) between the two LL transitions and the dominant absorption features of the zeroth-LL-related transitions, together with the DFT+DMFT-calculation-derived linear band dispersions in 2D momentum space, demonstrate that 2D MDF exist in the superconducting bulk state of NaFeAs. In addition, the LL transition energy of $\rm CaFe_2As_2$ displays the linear $\sqrt{B}$ dependence and the zero-energy intercept at $B = 0\rm ~T$. Moreover, the 2D-MDF Fermi velocities in NaFeAs, $\rm BaFe_2As_2$ and $\rm CaFe_2As_2$, which are extracted from the slopes of the linear $\sqrt{B}$ dependences of the LL transition energies, increase linearly with the Fe-As bond lengths—quantities about lattice degree of freedom. The linear scaling between the 2D-MDF Fermi velocities and the Fe-As bond lengths is supported by (i) the linear dependence of the $\sqrt{m^* } $ of the $d_{xy}$ electrons on the Fe-As bond length and (ii) the linear scaling between the $\sqrt{t_1 (xy,xy) + t_2 (xy,xy)}$ and the Fe-As bond length. Our wok offers a new material platform hosting both 2D MDF and superconductivity in the 3D bulk for discovering and tuning exotic novel quantum phenomena.

\textit{Acknowledgements.} We thank Zhiping Yin, Zhiqiang Li, Milan Orlita, Ying Ran, Oskar Vafek and Pierre Richard for very helpful discussions. The authors acknowledge support from the Guangdong Basic and Applied Basic Research Foundation (Projects No. 2021B1515130007), the strategic Priority Research Program of Chinese Academy of Sciences (Project No. XDB33000000), the National Natural Science Foundation of China (Grant No. U21A20432), the National Key Research and Development Program of China (Grant No. 2022YFA1403800), and the Synergetic Extreme Condition User Facility (SECUF, https://cstr.cn/31123.02.SECUF)-Infrared Unit in THz and Infrared Experimental Station. The single-crystal growth and characterization efforts at Rice are supported by the U.S. Department of Energy, Office of Basic Energy Sciences, under award no. DE-SC0012311 (P.D.). A portion of this work was performed in NHMFL which is supported by National Science Foundation Cooperative Agreement No. DMR-1157490 and the State of Florida.

$^{\dagger}$C.L., J.Z. and Y.X. contributed equally to this work. Z.-G.C. conceived and supervised this project. Y.S., C.Z. and P.D. grew the single crystals; N.-L.W. performed the infrared studies at zero magnetic field; Z.-G.C. carried out the magneto-infrared experiments; Partial magneto-infrared measurements were performed in NHMFL with the assistance from M.O.; J.Z. did the DFT+DMFT calculations; Z.-G.C., C.L. and Y.X. analyzed the data; Z.-G.C. wrote the paper with the input from C.L..

%\bibliography{NoEndingPoint}

\end{document}